\documentstyle[11pt,newpasp,twoside,epsf]{article}
\def\edcomment#1{\iffalse\marginpar{\raggedright\sl#1\/}\else\relax\fi}
\reversemarginpar
\newcommand{\xmm}{{\it XMM-Newton}}
\newcommand{\chandra}{{\it Chandra}}

\begin{document}
\title{Confronting atomic data with Fe L-shell spectra of stellar coronae}
 \author{Ehud Behar, Jean Cottam, John R. Peterson, Masao Sako, and Steven M. Kahn}
\affil{Columbia Astrophysics Laboratory, Columbia University, 550 West 120th Street,
New York, NY 10027-5247}

\author{Avraham Bar-Shalom, Marcel Klapisch}
\affil{ARTEP  inc. c/o Naval Research Laboratory, code 6730, 4555 Overlook Ave, SW Washington, DC 20375}
\author{Albert C. Brinkman}
\affil{Space Research Organization of the Netherlands, Sorbonnelaan 2, 3548 CA, Utrecht, The Netherlands}

\begin{abstract}
It is common to suggest the uncertainties in the atomic data for explaining difficulties that arise in interpreting astrophysical x-ray spectra. The atomic data for the Fe L-shell ions have, over the years, been considered particularly suspect. In this paper, we confront Fe-L spectra calculated using the HULLAC atomic code assuming coronal conditions with recent observations of the stellar coronae of Capella and HR 1099, obtained with the high-resolution grating spectrometers on board \chandra\ and \xmm. We find very good agreement, which indicates that Fe-L transition rates calculated with state-of-the-art codes, can generally be trusted in the analysis of x-ray spectra.
\end{abstract}

\section{Introduction}
Accurate understanding of atomic physics is crucial for the interpretation of spectroscopic, astrophysical observations. Uncertainties in atomic data can impede the ability to draw meaningful conclusions from observed spectra. Since a large amount of atomic data can be needed to interpret a single astrophysical spectrum, it is very difficult to predict how these uncertainties will affect the analysis; hence, the vast importance of reliable atomic data. With the advent of the high-resolution spectrometers on board \chandra\ and \xmm, the importance of the atomic physics relevant to the soft x-ray spectral regime has become even more pronounced.

The soft x-ray regime between 1 and 40 \AA\, which is covered complementally by the spectrometers on board \chandra\ and \xmm\, contains a rich forest of spectral lines emitted by many chemical elements at ionization stages corresponding to electron temperatures of $kT_{e}$ = 0.1 - 3 keV. At these high temperatures, most of the low-Z, cosmically abundant elements (C, N, O, Ne, Mg, Si, and S) are stripped down to their K shell (i.e., $n$ = 1, $n$ being the principal quantum number) and emit relatively few spectral lines. However, many strong lines of L-shell ($n$ = 2) Fe fall in this wavelength range as well. The eight ionization stages in the L shell can provide more precise information on the temperature structure of the source, independent of elemental abundances, than the two-ion K-shell systems. Brinkman et al. (2001) and Peterson et al. (2001), for instance, have recently used this quality of the Fe-L spectra. In general, transitions for both K- and L- shell ions can be calculated with available state-of-the-art atomic codes. Nevertheless, the Fe L-shell lines have been considered less certain (presumably because of the increasing complexity of multi-electron atoms). A few examples can be found in Kaspi et al. (2000), Harrus et al. (1997), and Masai (1997). 

The atomic processes associated with the emission from an astrophysical source depend on the type of plasma at hand. In this work, we focus on low-density steady-state coronal plasmas that are dominated by electron-ion collisional processes and subsequent spontaneous radiative decays. Stellar coronae ought to be especially simple in terms of the atomic-state kinetics, similar to tokamaks and ion traps in the laboratory. By comparing our calculations with the spectra of the bright stellar coronal sources Capella and HR 1099, we aim to confirm that just like the K-shell atomic data, the Fe L-shell data, when calculated correctly, are highly reliable.

\section{The Atomic-State Kinetic Model}
It is useful to distinguish between atomic processes within charge states and those from one charge state to the other (i.e., ionization and recombination), because the rates for the latter can be much harder to measure or to calculate. Particularly challenging are excitation autoionization and dielectronic recombination (DR) processes, which can involve many thousands of doubly-excited intermediate levels. Thus, we advocate a method for modeling in which each ion is treated separately. This approximation is appropriate in cases where the ionization and recombination processes do not produce lines directly. Steady state, coronal plasmas tend to fall into this category. In this model, the contribution of each ion to the spectrum is scaled separately to match one of the strongest lines of that ion in the observed spectrum. This method is quite different from the commonly used (e.g., Audard et al. 2001) global fitting methods. The avoidance of global fitting simplifies the procedure considerably and allows for a direct test of the atomic calculations, without the various statistical biases imposed by global fitting. Each ion is assumed to emit at one electron temperature. This is a reasonable approximation for most L-shell ions, which form in a relatively narrow temperature range. In this range, the spectrum within each ionization state has only a moderate dependence on the temperature. The lines and the continuum in our model are not treated on an equal footing, since continuum processes, unlike discrete line emission, do not have unique signatures in the spectrum. Even marginal uncertainties in the line profiles (especially the broad wings) in the line-crowded Fe-L regions of the spectrum can confuse the evaluation of the continuum. However, the errors in estimating the continuum have a very small effect on measured line ratios, which are the ultimate tool of spectroscopic analysis.

In stellar coronal sources, all the ions are generally in the ground level, with the exception of metastable states. Spectral x-ray lines are produced when a plasma electron collisionally excites an ion that subsequently decays radiatively, either directly to the ground level or via cascades. Hence, the power of a given line depends on the rate coefficients for collisional excitation from the ground level and the rates for radiative decay from excited levels. We have carried out {\it ab-initio} level-by-level calculations for each of the Fe L-shell ions from Ne-like Fe$^{16+}$ through Li-like Fe$^{23+}$. All of the collisional and radiative transitions within each ionization state are included. In order to account for the effect of radiative cascades, high lying levels with $n$-values as high as 12, but at least 6 are included in the model. 

The collisional and radiative rates are obtained using the multi-configuration, relativistic HULLAC (Hebrew University Lawrence Livermore Atomic Code) computer package (Bar-Shalom, Klapisch, \& Oreg 2001). The intermediate-coupling level energies are calculated using the relativistic version (Koenig 1972; Klapisch et al. 1977) of the Parametric Potential method by Klapisch (1971). The collisional rate coefficients are calculated in the distorted wave approximation, implementing the highly efficient factorization-interpolation method by Bar-Shalom, Klapisch, \& Oreg (1988). The angular momentum algebra is computed very efficiently using graphical methods (Bar-Shalom \& Klapisch 1988).

\section{Confronting the Atomic Data with Observations}
Using the very simple single-ion model described above, we test the accuracy of the HULLAC atomic data by comparison with the x-ray spectra of the stellar coronal sources Capella and HR 1099 measured, respectively, with the High Energy Transmission Grating Spectrometer (HETGS) on board \chandra\ and the Reflection Grating Spectrometer (RGS) on board \xmm. Both sources show a line rich spectrum with relatively low continuum. A more comprehensive presentation of these observations, including their interesting astrophysical interpretation can be found in Behar, Cottam, \& Kahn (2001) and in Brinkman et al. (2001), respectively. In the present work, we limit ourselves to testing the quality of the HULLAC data by taking a closer look at some of the weak Fe L-shell spectral features and how well they are reproduced by the model.

\subsection{Capella with the HETGS}
The fluxed spectrum of Capella and the present model are those used in Behar et al. (2001). A comparison of the resulting positive and negative first order fluxes for the Medium Energy Grating (MEG) shows an agreement of 20\% (or better) from 7 to 18 \AA\, which should give an idea of the uncertainties in the measurement. The background and continuum in the spectrum are very low and can be neglected in the model. The high spectral resolution of HETGS enables a comparison between the observed and calculated spectra almost line by line. In Behar et al. (2001), it was shown that the Fe-L HETGS spectrum of Capella can be reproduced almost entirely by assuming a single electron temperature of  $kT_{e}$ = 0.6 keV for Fe$^{16+}$ through Fe$^{21+}$. In order to show how well the model compares with the observed spectrum, in Fig. 1, we compare the data and the Fe model in the 10.90 - 12.35 \AA\ range. This range is chosen, because it contains relatively weak lines emitted by ions that also have stronger lines. Since the individual-ion contributions are determined by the latter, the accuracy of the former is a test to the quality of the atomic calculations. To prevent confusion, the non-Fe blends are not included in the theoretical spectrum, but nonetheless are indicated explicitly in the figure. The theoretical spectral lines in Fig. 1 are given a uniform Gaussian profile with 15 m\AA\ full width at half maximum (FWHM). The HULLAC wavelengths are usually only accurate to within 10 - 20 m\AA. For a line-by-line comparison see Behar et al. (2001). These inaccuracies do not impede the line identifications and in Fig. 1 even come in handy, facilitating the presentation of the comparison.

\begin{figure}
\plotone{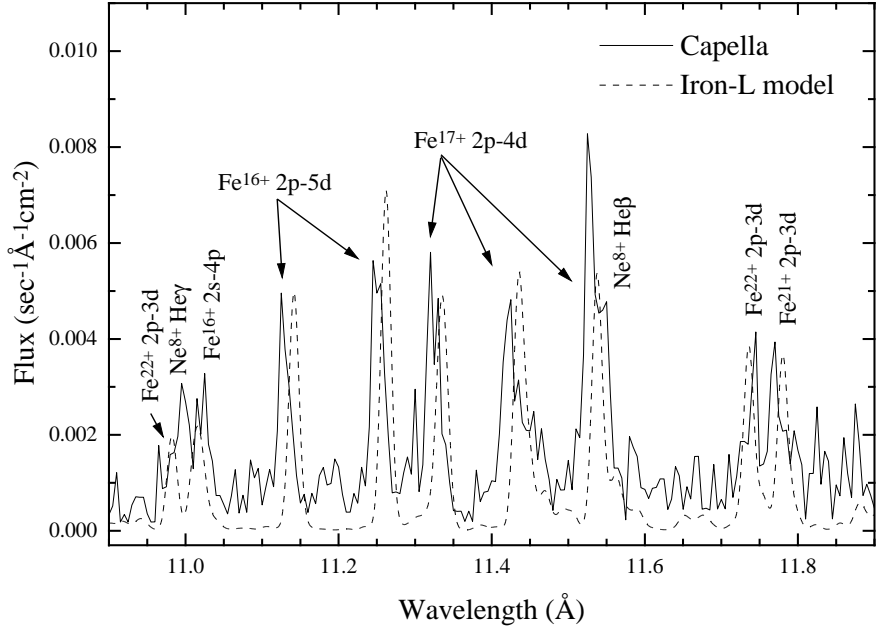}
\caption{Comparison of the observed MEG Capella spectrum with the present Fe (only) model based on HULLAC atomic data. The strongest lines in this limited range are identified.}
\end{figure}

It can be seen that the observed line intensities are reproduced by the model usually to within 20\%, which is about the uncertainty of the measured intensities. In particular, see the 2p - 5d lines of Fe$^{16+}$ and the 2p - 4d lines of Fe$^{17+}$. Note that the theoretical intensities are normalized to the strong 2p - 3d lines of these ions at longer wavelengths (not in the figure). Thus, the good agreement of the intensities for these rather weak lines is a direct result of the accuracy of the relevant atomic data. Larger discrepancies occur for the Fe$^{16+}$ 2s-4p line at 11.022 \AA\ and for the Fe$^{17+}$ 2p - 4d line at 11.527 \AA\, but these can be at least partially attributed to the blending with Ne$^{8+}$ lines that are not included in the model. Assessing the effect of blending more precisely requires exact wavelengths and accurate instrumental line profiles. Additionally, the data presented here for Capella pertain to a rather short observation (15 ksec). Spectra obtained by superimposing many observations, as done by Canizares et al. (2000) for Capella or as done below for HR 1099 provide more stringent tests to the atomic rates.
\begin{figure}
\plotone{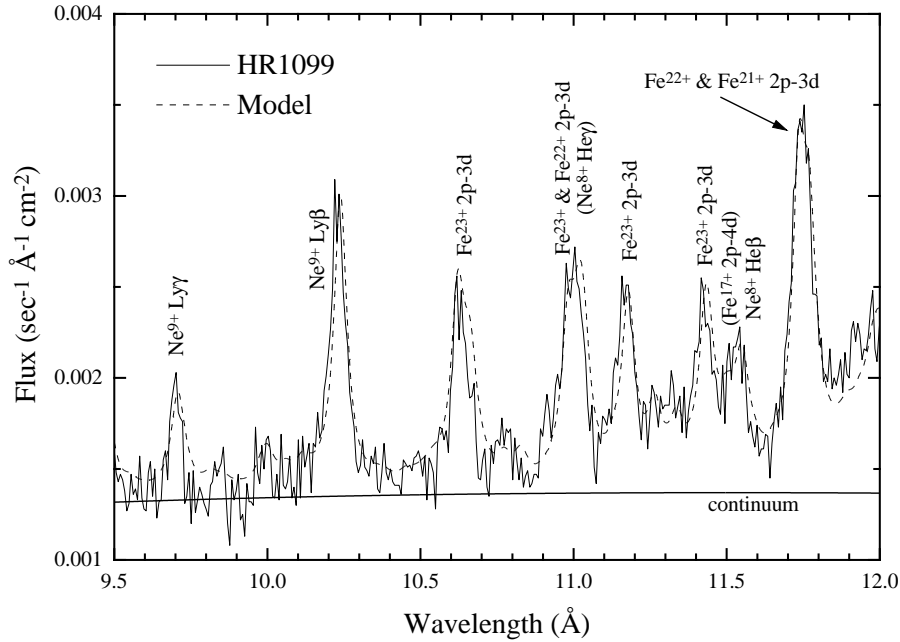}
\caption{Comparison of the observed RGS HR 1099 spectrum with the present model based on HULLAC atomic data. The strongest lines in this limited range are labeled. Note the very good agreement both for the Ne-K and for the Fe-L lines.}
\end{figure}

\subsection{HR 1099 with the RGS}
The spectrum and model used here for HR 1099 are those we presented in the RGS first light paper (Brinkman et al. 2001). As opposed to Capella, for HR 1099 all eight L-shell ionization states of Fe can be clearly detected in the spectrum. This is indicative of a broad distribution of temperatures. Therefore, for HR 1099 we assume that each ion emits mostly around the electron temperature at which it has its maximum abundance. For HR 1099, the continuum cannot be neglected and it is estimated separately by a phenomenological bremsstrahlung component. In Fig. 2, we focus on a limited region (9.5 - 12.0 \AA) in the HR 1099 spectrum that includes many Fe-L lines. In this case, the non-Fe elements are also included in the model for the sake of completeness. For convenience, the theoretical spectral lines in Fig. 2 are given Lorentzian profiles with a width that slightly varies with wavelength, as prescribed by the RGS instrumental model. These profiles match the observed profiles fairly well although minor discrepancies remain in the broad wings of the profile due to the known instrumental scattering wings, which are not picked up by the Lorentzian. Refinements in the calibration of the RGS line profiles are still in progress (den Herder et al. 2001). In order to better account for the effects of line blends, this time we have corrected the Fe-L HULLAC wavelengths according to the precise laboratory measurements by Brown et al. (2000).

In Fig. 2, by looking at the peak heights, it can be seen that the observed Fe-L line intensities, just like the Ne-K intensities, are reproduced by the model to within a few percent. The better agreement for HR 1099 is mostly due to the very long (570 ksec) observation, which includes approximately 1.3 million total source counts in the 5 - 35 \AA\ band. Many of the Fe-L lines shown in Fig. 2 are 2p-3d lines of Fe$^{22+}$ and Fe$^{23+}$, but they are blended with transitions from high-$n$ levels in the lower Fe-L charge states, which are clearly seen in Capella, where the higher charge states are absent. Note that each ionization state introduces only one degree of freedom required to fit many lines pertaining to that ionization state. Also, the HULLAC code is complete in producing all of the lines emitted by each charge state and taking all of the significant cascade effects into account. In short, the generally good agreement between the data and model, found for the entire Fe-L spectral range, is a manifestation of the adequacy of the atomic rates. 
\section{Afterthoughts}
The natural question arises, whether these observations by \chandra\ and \xmm\ are really providing the first opportunity to benchmark the validity of these atomic calculations. An honest answer would have to be no. Two useful sources for x-ray spectra of hot collisional gas are solar flares (e.g., McKenzie et al. 1980) and the electron beam ion trap (EBIT) (Levine et al. 1988) that have been available for some time now. The parts of the HULLAC code needed for the present calculations were available since 1988. The fundamental atomic physics needed for computing the present data is well known and the current HULLAC rates could have been easily calculated as early as ten years ago and compared with high-resolution solar or laboratory measurements. Recently, Brown et al. (1998) compared Fe$^{16+}$ intensity ratios measured with EBIT to test ratios calculated by HULLAC (see Figs. 4, 5, and 6 therein) and found very good agreement. Gu et al. (1998) compared cross sections for producing Fe$^{23+}$ line emission (including cascades) and found that, away from threshold, the calculated cross sections are accurate to within a few percent. Fortunately, the atomic data used here are not unique to the present model. The latest versions of the widely used spectral fitting codes SPEX (Kaastra, Mewe, \& Nieuwenhuijzen 1996) and XSPEC (Arnaud 1996) now incorporate similar data (from an earlier version of HULLAC) in their respective databases: MEKAL (Mewe, Kaastra, \& Liedahl 1995) and APEC (Smith et al. 1999), although updates are still in progress. Comparisons of the present model with APEC show good agreement for the Fe-L collisionally excited lines. The current APEC version still has DR satellite lines erroneously concentrated at one wavelength for each ion, but these will be removed soon (Smith 2000). Thus, those codes should soon provide their users with data as adequate as the data used here. If for nothing else, we should thank the \chandra\ and \xmm\ missions for calling attention to this cornerstone of our research.

Despite the ability to unambiguously interpret almost all of the soft x-ray coronal spectra, this is not to say that all of the atomic physics is completely understood. For example, the role of resonant excitations (i.e., dielectronic capture followed by autoionization to excited levels) in producing the 2p-3s lines of Fe$^{17+}$ around 16 \AA\ has yet to be investigated (see Behar et al. (2001) for the discrepancies between the model and the data for these lines). Furthermore, although we would expect the accuracy of HULLAC calculations for the lines of L-shell Si, S, Ar, and Ca that are in the RGS wavelength band to be comparable to that of Fe-L, a thorough investigation of these ions has not yet been performed. 
\section{Conclusions}
We have shown that a very simple coronal model, which uses HULLAC collisional and radiative transition rates for the Fe L-shell ions, reproduces well the high-resolution x-ray spectra of Capella and HR 1099 as measured with the grating spectrometers on board \chandra\ and \xmm. This is interpreted as a verification of the rates for the collisional and radiative atomic processes within charge states in hot, low-density, collisional plasmas. In particular, there is no reason to believe that the data for the (Fe) L-shell ions should be less reliable than the K-shell data. 

\acknowledgments
We are grateful to the RGS team for collaboration throughout this work. We acknowledge generous support from the National Aeronautics and Space Administration. J.C. was funded by the NASA GSRP program under grant number NGT 5-50152.

\end{document}